\documentclass[runningheads]{llncs}

\usepackage{eccv}

\usepackage{eccvabbrv}

\usepackage{graphicx}
\usepackage{booktabs}

\usepackage[accsupp]{axessibility}  

\usepackage[pagebackref,breaklinks,colorlinks,citecolor=eccvblue]{hyperref}

\usepackage{orcidlink}

\newsavebox\CBox
\def\textBF#1{\sbox\CBox{#1}\resizebox{\wd\CBox}{\ht\CBox}{\textbf{#1}}}

\begin{document}

\title{NCT-CRC-HE: Not All Histopathological\\ Datasets Are Equally Useful}


\author{Andrey Ignatov\inst{1,2,*} \and
Grigory Malivenko\inst{2,*}}

\authorrunning{Andrey Ignatov, Grigory Malivenko}

\institute{Computer Vision Lab, ETH Zurich, Switzerland \and
AI Witchlabs Ltd, Switzerland \\
$^*$ Indicates equal contribution \\
\email{andrey@vision.ee.ethz.ch, grigory.malivenko@gmail.com}}

\maketitle

\begin{abstract}
  Numerous deep learning-based solutions have been proposed for histopathological image analysis over the past years. While they usually demonstrate exceptionally high accuracy, one key question is whether their precision might be affected by low-level image properties not related to histopathology but caused by microscopy image handling and pre-processing. In this paper, we analyze a popular NCT-CRC-HE-100K colorectal cancer dataset used in numerous prior works and show that both this dataset and the obtained results may be affected by data-specific biases. The most prominent revealed dataset issues are inappropriate color normalization, severe JPEG artifacts inconsistent between different classes, and completely corrupted tissue samples resulting from incorrect image dynamic range handling. We show that even the simplest model using only 3 features per image (red, green and blue color intensities) can demonstrate over 50\% accuracy on this 9-class dataset, while using color histogram not explicitly capturing cell morphology features yields over 82\% accuracy. Moreover, we show that a basic EfficientNet-B0 ImageNet pretrained model can achieve over 97.7\% accuracy on this dataset, outperforming all previously proposed solutions developed for this task, including dedicated foundation histopathological models and large cell morphology-aware neural networks. The NCT-CRC-HE dataset is publicly available and can be freely used to replicate the presented results. The codes and pre-trained models used in this paper are available at \url{https://github.com/gmalivenko/NCT-CRC-HE-experiments}.

\end{abstract}

\section{Introduction}
\label{sec:intro}

Digital histopathology is a rapidly evolving field that focuses on automatic computer-assisted analysis of high-resolution microscopy photos of stained tissue regions, also called whole slide images (WSIs). These tissue photos provide lots of valuable morphological information on the cellular level that is relevant for clinical diagnostics, including cell type composition and cell-cell interactions, activity of the immune system, cell cycle progression, various abnormalities in cell structure and shape that are often good indicators of cellular stress, \textit{etc}. Previous research works demonstrated that this histopathological data can be used for designing diagnostic tools for many different biomedical tasks including tissue lesion detection and cancer classification~\cite{hou2016patch,bandi2018detection,tsaku2019texture,song2020clinically,yu2021large,komura2022universal,kumar2023crccn,kang2023benchmarking,anju2023finetuned,lu2023towards,ignatov2024histopathological}, tumor grading~\cite{raju2020graph,koziarski2021diagset,xu2021computerized,barbano2021unitopatho,lomenie2022can,bulten2022artificial}, predicting gene mutants~\cite{coudray2018classification,yamashita2021deep,qu2021genetic}, biomarkers~\cite{lu2022slidegraph+,wagner2023transformer} and overall gene expression levels~\cite{schmauch2020deep,dawood2021all}, detecting mitosis~\cite{wang2014mitosis,balkenhol2019deep,nateghi2021deep}, quantifying the activity of the immune system~\cite{turkki2016antibody,xu2021deep,abousamra2022deep}, predicting patient survival \cite{fu2020pan,wulczyn2020deep,yao2020whole,agarwal2021survival,shao2021weakly,tsai2023histopathology}, \textit{etc}.

A large amount of rich visual data provided by WSIs led to a rapid development of various deep learning-based solutions for the analysis of histopathological images. As deep neural networks can automatically learn complex patterns directly from the data, taking into account all morphological features and revealing hidden data structures, they were able to achieve top results on the majority of whole slide image analysis tasks~\cite{wang2022transformer,chen2024towards,ignatov2024histopathological,wagner2023transformer,wang2023retccl}, often outperforming the results demonstrated by professional pathologists. However, the real predictive power of such solutions strongly depends on the quality of the datasets used for their training, and might be biased towards some specific data properties not related to the task itself. When it comes to histopathological datasets, the biggest source of bias here is related to the overall data formation procedure: as one usually cannot collect data for multiple diseases or patients in the same institution, large-scale datasets represent a compilation of microscopy images obtained in different laboratories or even countries. This often leads to a pronounced batch effect: since images are collected with different equipment, by different technicians using slightly varying tissue staining / handling techniques, and additionally post-processed with different libraries and tools, they might contain site-specific signatures that can be used to uniquely identify image origin~\cite{howard2021impact}. While this variation might not be an issue when all images are sampled randomly from different places, in practice each laboratory usually specializes in a specific disease or tissue type, and thus the entire data for some classes is often obtained in one specific place, encompassing the corresponding low-level image signatures. A number of image normalization methods have been proposed to deal with this issue~\cite{reinhard2001color,macenko2009method,vahadane2016structure,zheng2019adaptive,kang2021stainnet}, however, several research works indicate low efficiency of such tools in eliminating all inherent site-specific image properties~\cite{gupta2017automated,tellez2019quantifying,voon2023evaluating}. Therefore, one key question remains: do the advanced deep learning methods form their decision rules based on disease-specific tissue morphology, or they largely rely on variation in staining, resolution and image processing artifacts specific for each tissue class.

\begin{figure}[t!]
\centering
\setlength{\tabcolsep}{1pt}
\resizebox{1.0\linewidth}{!}
{
\begin{tabular}{ccccccccc}
 & & & & & & Normal Colon & Cancer-Associated & Colorectal \\
Adipose & Background & Debris & Lymphocyte & Mucus & Smooth Muscle & Mucosa & Stroma & Adenocarcinoma \\
\includegraphics[width=0.25\linewidth]{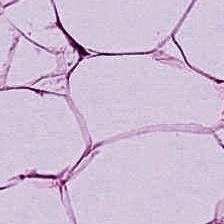}&
\includegraphics[width=0.25\linewidth]{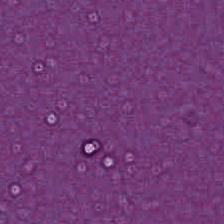}&
\includegraphics[width=0.25\linewidth]{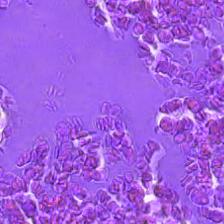}&
\includegraphics[width=0.25\linewidth]{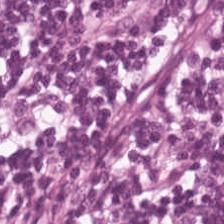}&
\includegraphics[width=0.25\linewidth]{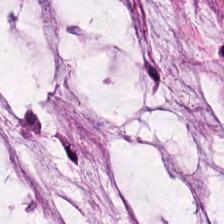}&
\includegraphics[width=0.25\linewidth]{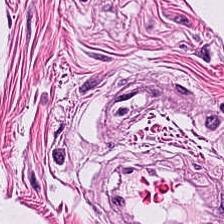}&
\includegraphics[width=0.25\linewidth]{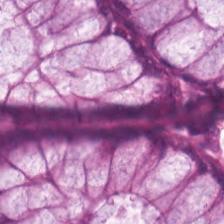}&
\includegraphics[width=0.25\linewidth]{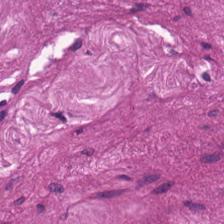}&
\includegraphics[width=0.25\linewidth]{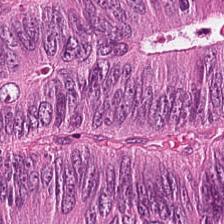}
\\
\includegraphics[width=0.25\linewidth]{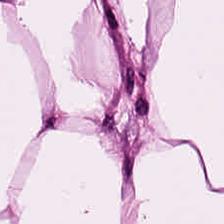}&
\includegraphics[width=0.25\linewidth]{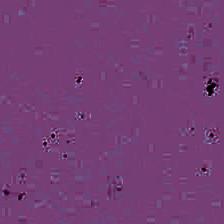}&
\includegraphics[width=0.25\linewidth]{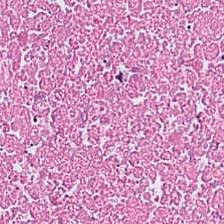}&
\includegraphics[width=0.25\linewidth]{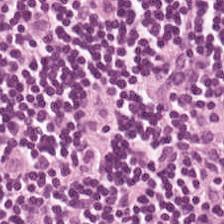}&
\includegraphics[width=0.25\linewidth]{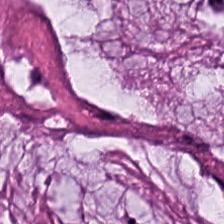}&
\includegraphics[width=0.25\linewidth]{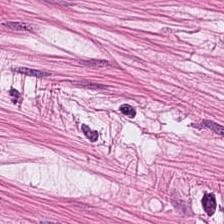}&
\includegraphics[width=0.25\linewidth]{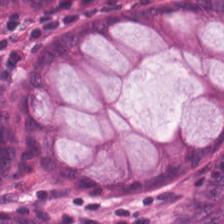}&
\includegraphics[width=0.25\linewidth]{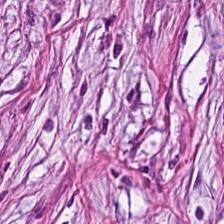}&
\includegraphics[width=0.25\linewidth]{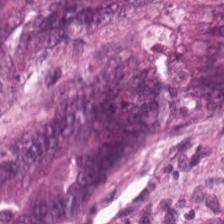}
\\
\includegraphics[width=0.25\linewidth]{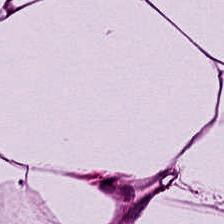}&
\includegraphics[width=0.25\linewidth]{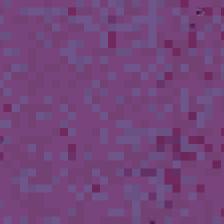}&
\includegraphics[width=0.25\linewidth]{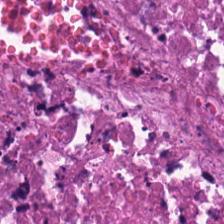}&
\includegraphics[width=0.25\linewidth]{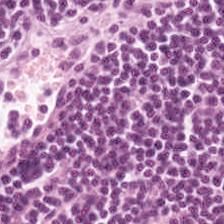}&
\includegraphics[width=0.25\linewidth]{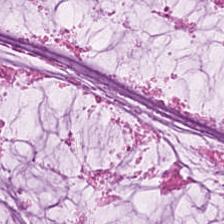}&
\includegraphics[width=0.25\linewidth]{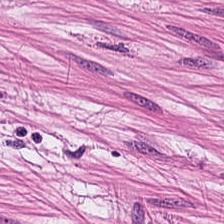}&
\includegraphics[width=0.25\linewidth]{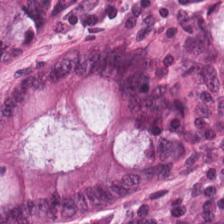}&
\includegraphics[width=0.25\linewidth]{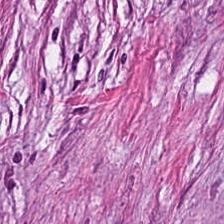}&
\includegraphics[width=0.25\linewidth]{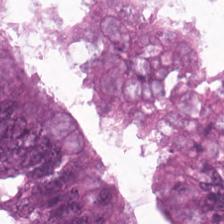}
\\
\includegraphics[width=0.25\linewidth]{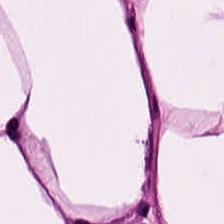}&
\includegraphics[width=0.25\linewidth]{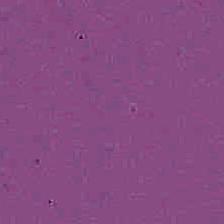}&
\includegraphics[width=0.25\linewidth]{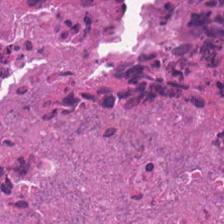}&
\includegraphics[width=0.25\linewidth]{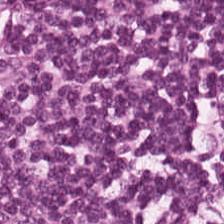}&
\includegraphics[width=0.25\linewidth]{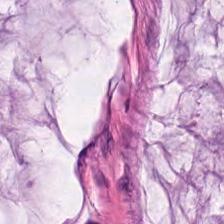}&
\includegraphics[width=0.25\linewidth]{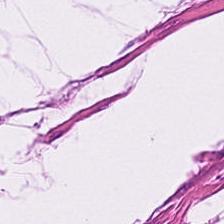}&
\includegraphics[width=0.25\linewidth]{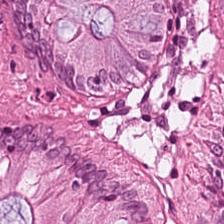}&
\includegraphics[width=0.25\linewidth]{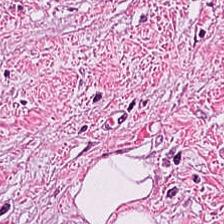}&
\includegraphics[width=0.25\linewidth]{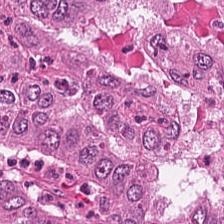}
\\
\includegraphics[width=0.25\linewidth]{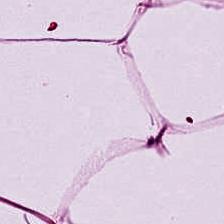}&
\includegraphics[width=0.25\linewidth]{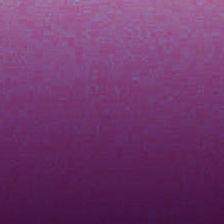}&
\includegraphics[width=0.25\linewidth]{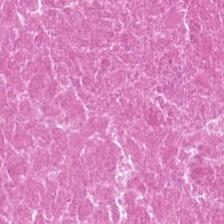}&
\includegraphics[width=0.25\linewidth]{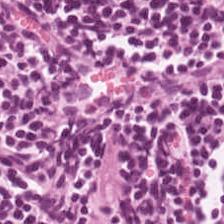}&
\includegraphics[width=0.25\linewidth]{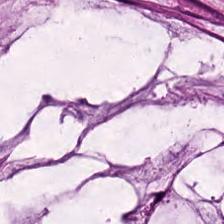}&
\includegraphics[width=0.25\linewidth]{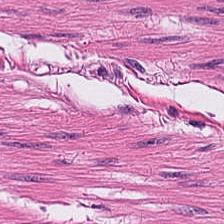}&
\includegraphics[width=0.25\linewidth]{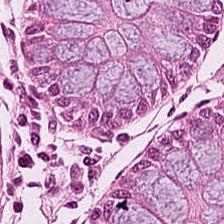}&
\includegraphics[width=0.25\linewidth]{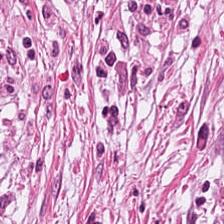}&
\includegraphics[width=0.25\linewidth]{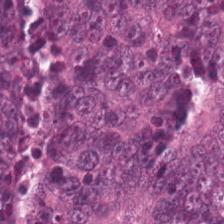}
\\
\includegraphics[width=0.25\linewidth]{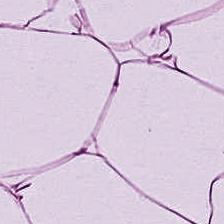}&
\includegraphics[width=0.25\linewidth]{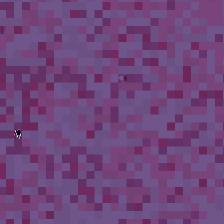}&
\includegraphics[width=0.25\linewidth]{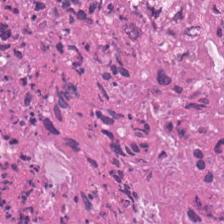}&
\includegraphics[width=0.25\linewidth]{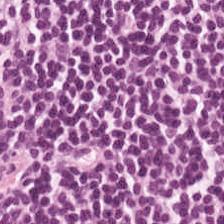}&
\includegraphics[width=0.25\linewidth]{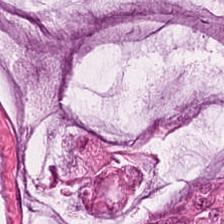}&
\includegraphics[width=0.25\linewidth]{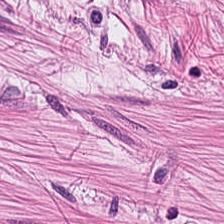}&
\includegraphics[width=0.25\linewidth]{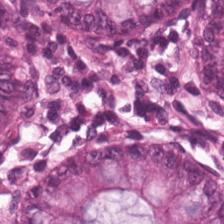}&
\includegraphics[width=0.25\linewidth]{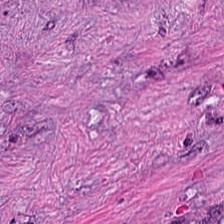}&
\includegraphics[width=0.25\linewidth]{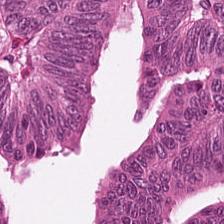}
\\
\includegraphics[width=0.25\linewidth]{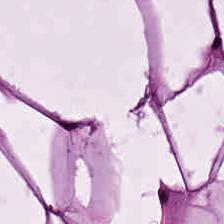}&
\includegraphics[width=0.25\linewidth]{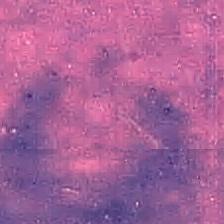}&
\includegraphics[width=0.25\linewidth]{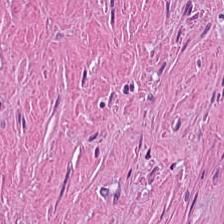}&
\includegraphics[width=0.25\linewidth]{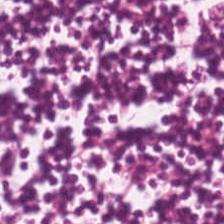}&
\includegraphics[width=0.25\linewidth]{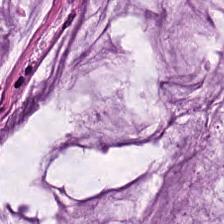}&
\includegraphics[width=0.25\linewidth]{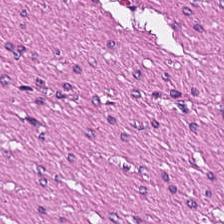}&
\includegraphics[width=0.25\linewidth]{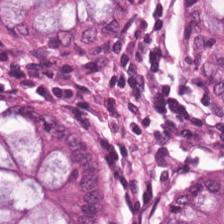}&
\includegraphics[width=0.25\linewidth]{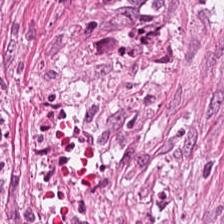}&
\includegraphics[width=0.25\linewidth]{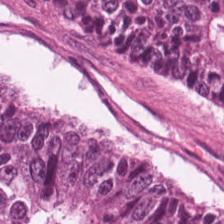}
\\
\includegraphics[width=0.25\linewidth]{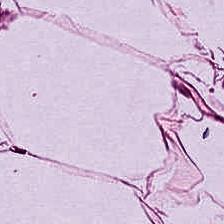}&
\includegraphics[width=0.25\linewidth]{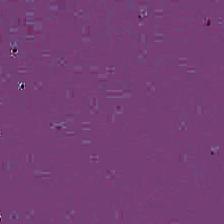}&
\includegraphics[width=0.25\linewidth]{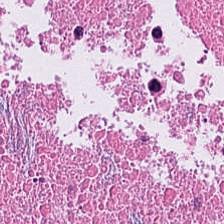}&
\includegraphics[width=0.25\linewidth]{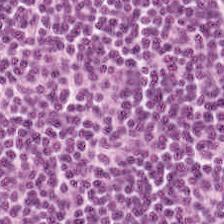}&
\includegraphics[width=0.25\linewidth]{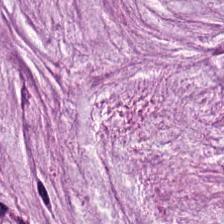}&
\includegraphics[width=0.25\linewidth]{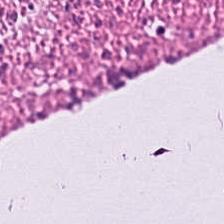}&
\includegraphics[width=0.25\linewidth]{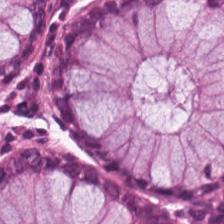}&
\includegraphics[width=0.25\linewidth]{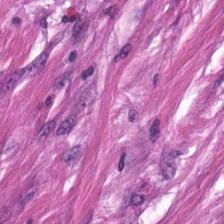}&
\includegraphics[width=0.25\linewidth]{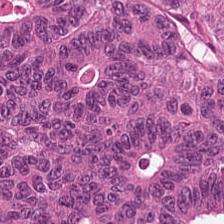}
\\
\includegraphics[width=0.25\linewidth]{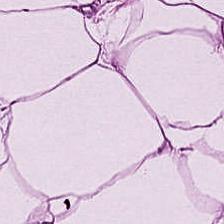}&
\includegraphics[width=0.25\linewidth]{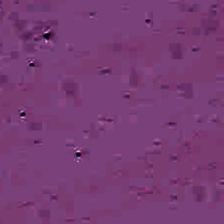}&
\includegraphics[width=0.25\linewidth]{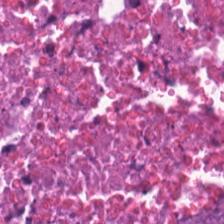}&
\includegraphics[width=0.25\linewidth]{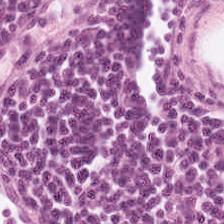}&
\includegraphics[width=0.25\linewidth]{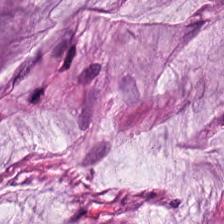}&
\includegraphics[width=0.25\linewidth]{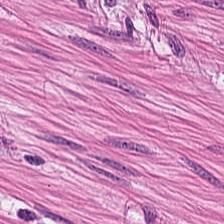}&
\includegraphics[width=0.25\linewidth]{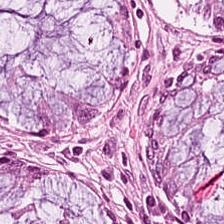}&
\includegraphics[width=0.25\linewidth]{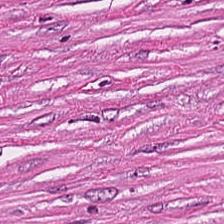}&
\includegraphics[width=0.25\linewidth]{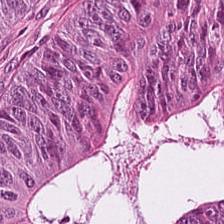}
\\
\includegraphics[width=0.25\linewidth]{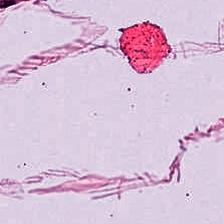}&
\includegraphics[width=0.25\linewidth]{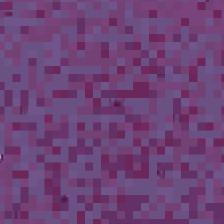}&
\includegraphics[width=0.25\linewidth]{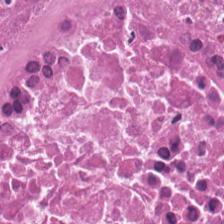}&
\includegraphics[width=0.25\linewidth]{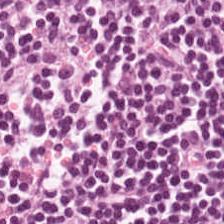}&
\includegraphics[width=0.25\linewidth]{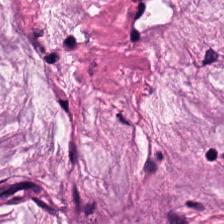}&
\includegraphics[width=0.25\linewidth]{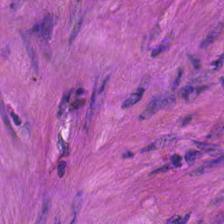}&
\includegraphics[width=0.25\linewidth]{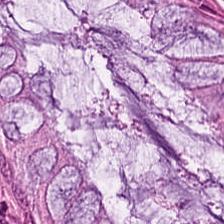}&
\includegraphics[width=0.25\linewidth]{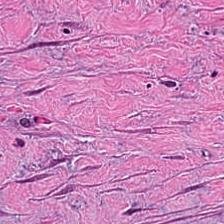}&
\includegraphics[width=0.25\linewidth]{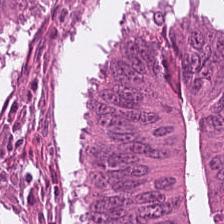}
\\
\end{tabular}
}
\caption{Visualization of normalized H\&E stained image patches from the NCT-CRC-HE-100K dataset. The images were sampled randomly for each of 9 tissue classes.}
\vspace{-2mm}
\label{fig:NCT_sample_images}
\end{figure}

In this work, we focus on the exploration of the NCT-CRC-HE~\cite{kather2019predicting} colorectal cancer dataset consisting of 100,000 training / 7,180 test image patches belonging to nine tissue classes: \textit{adipose, background, debris, lymphocyte, mucus, smooth muscle, normal colon mucosa, cancer-associated stroma} and \textit{colorectal adenocarcinoma epithelium}. This dataset gained high popularity among the research community with numerous approaches proposed for tissue classification and patient survival prediction, starting from basic CNNs~\cite{khvostikov2021tissue,kather2019predicting,wang2021accurate,sun2023automatic,anju2023finetuned} to advanced foundation transformer models~\cite{lu2023towards,filiot2023scaling,kang2023benchmarking,wang2022transformer} and dedicated cell morphology-aware networks~\cite{ignatov2024histopathological}. Besides the large size, one of the main advantages of the NCT-CRC-HE dataset is its fixed test set containing data from 50 independent patients, which should potentially remove some bias. However, different inconsistencies in the results reported on this dataset and atypical learning curves obtained during model training suggested potential issues with the data. A brief subsequent visual analysis of real training and validation data (Fig.~\ref{fig:NCT_sample_images}) confirmed the initial concerns, showing various image pre-processing issues explaining the observed results and model behavior.

This paper provides an overview of the NCT-CRC-HE training and test sets, analyzing various found inconsistencies and their potential effect on the final deep learning models and their results. In particular, we demonstrate that there exists a strong color signature for the majority of tissue classes that allows to correctly classify more than half of the test images by using only 3 features per each image~--- red, green and blue average color intensities. Switching to a basic color histogram encoding the variations in tissue staining leads to correct classification of 8 out of 10 images without using any deep learning models. Besides that, we show that some tissue classes suffer from strong JPEG compression artifacts, which are easily identifiable even by simplest CNN models and can be used on their own for unique image identification. Another issue is related to corruptions presumably caused by incorrect image dynamic range handling that results in patches that no long have any biological meaning. Finally, we show that by taking into account the above mentioned issues and training a tiny EfficientNet-B0 model on this data, one can achieve the state-of-the-art accuracy of 97.7\%, outperforming all previously proposed dedicated solutions developed for the considered dataset. This suggests that no advanced histopathology-related features are needed to correctly classify images from the CRC-VAL-HE-7K test set, and this should be taken into account when designing and interpreting all future results obtained on this dataset.

\section{Exploring and Analyzing the NCT-CRC-HE Dataset}
\label{sec:nct_exploration}

NCT-CRC-HE dataset~\cite{kather2019predicting} consists of two independent partitions: \textit{NCT-CRC-HE-100K} with 100,000 training patches extracted from 86 whole slide images, and \textit{CRC-VAL-HE-7K} containing 7180 test patches from 50 separate patients with colorectal adenocarcinoma. The corresponding tissue samples combine data obtained from the tissue bank of the National Center for Tumor diseases (NCT) and the pathology archive at the University Medical Center Mannheim (UMM). All images were normalized with the Macenko method~\cite{macenko2009method}, the resolution of the extracted patches is 224$\times$224 pixels. The dataset is publicly available and can be downloaded from \url{https://zenodo.org/records/1214456}.

The initial visual inspection of patches belonging to different tissue classes (Fig.~\ref{fig:NCT_sample_images}) indicated the presence of various artifacts on the considered images and a potential difference in color intensities for different tissue classes. Therefore, a more detailed analysis of the found issues was performed to analyze their severity and potential effect on the trained deep learning models.

\subsection{RGB Channel Intensities and Color Distribution}

When observing visualized image crops (Fig.~\ref{fig:NCT_sample_images}), one can notice the difference in the color intensity / brightness for different tissue classes. In principle, this difference should be partly eliminated by using various stain normalization techniques~\cite{reinhard2001color,macenko2009method,vahadane2016structure,zheng2019adaptive,kang2021stainnet} developed to reduce any potential batch effect. The authors of the NCT-CRC-HE dataset used the Macenko normalization method~\cite{macenko2009method}, nevertheless, the normalized images still have a pronounced color signature.

\begin{figure*}[h!]
  \centering
   \includegraphics[width=0.32\linewidth]{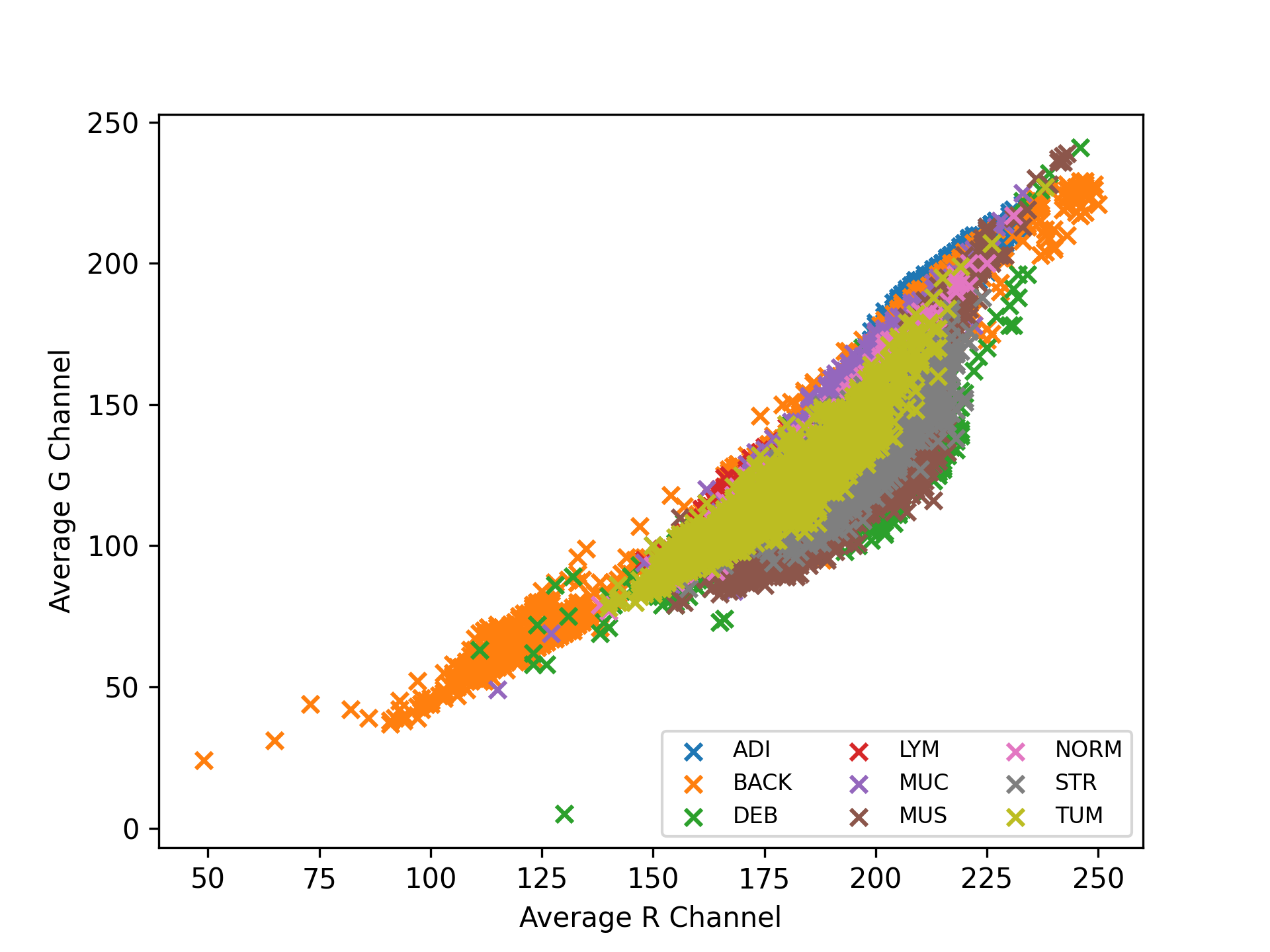}
   \includegraphics[width=0.32\linewidth]{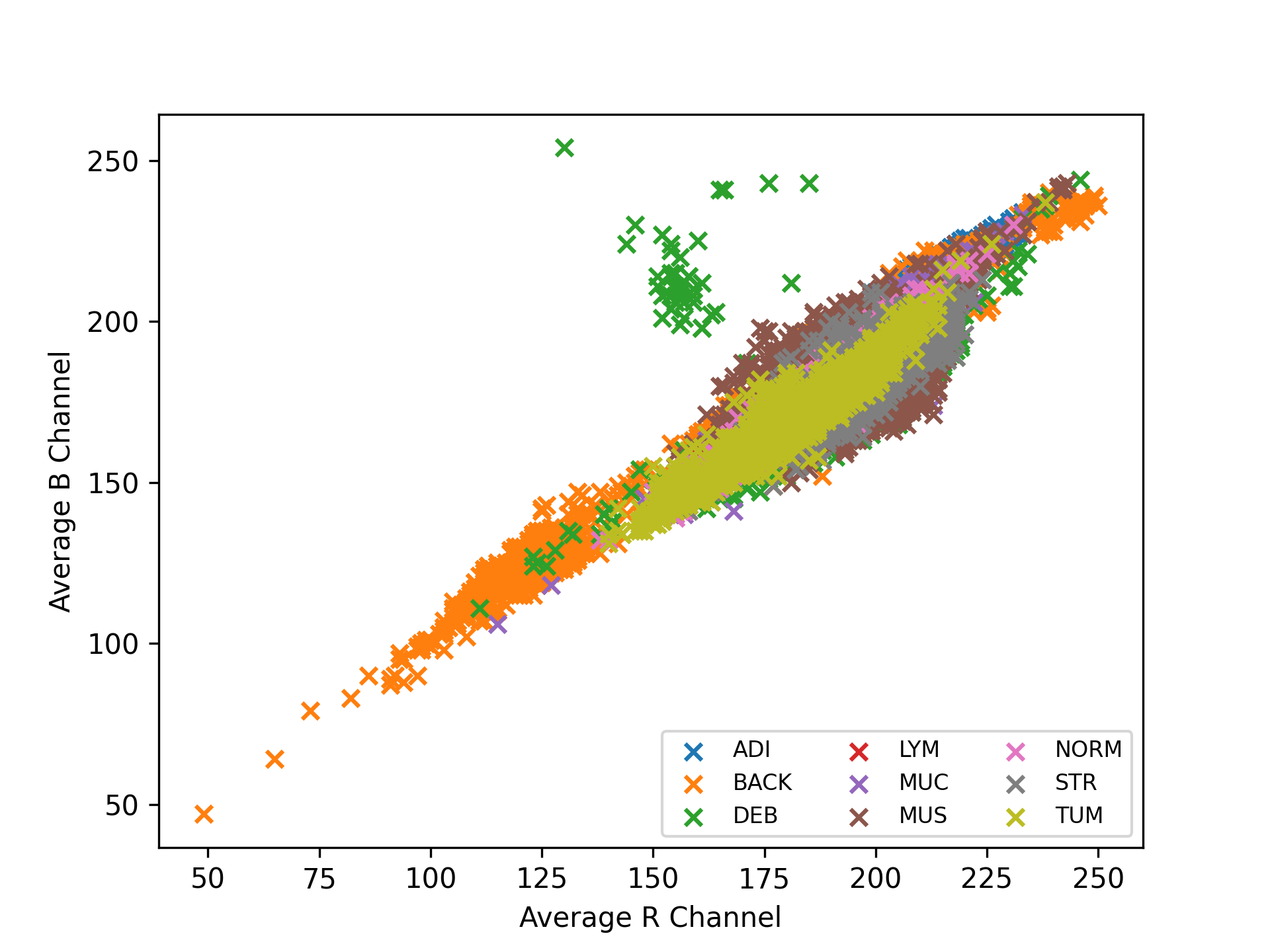}
   \includegraphics[width=0.32\linewidth]{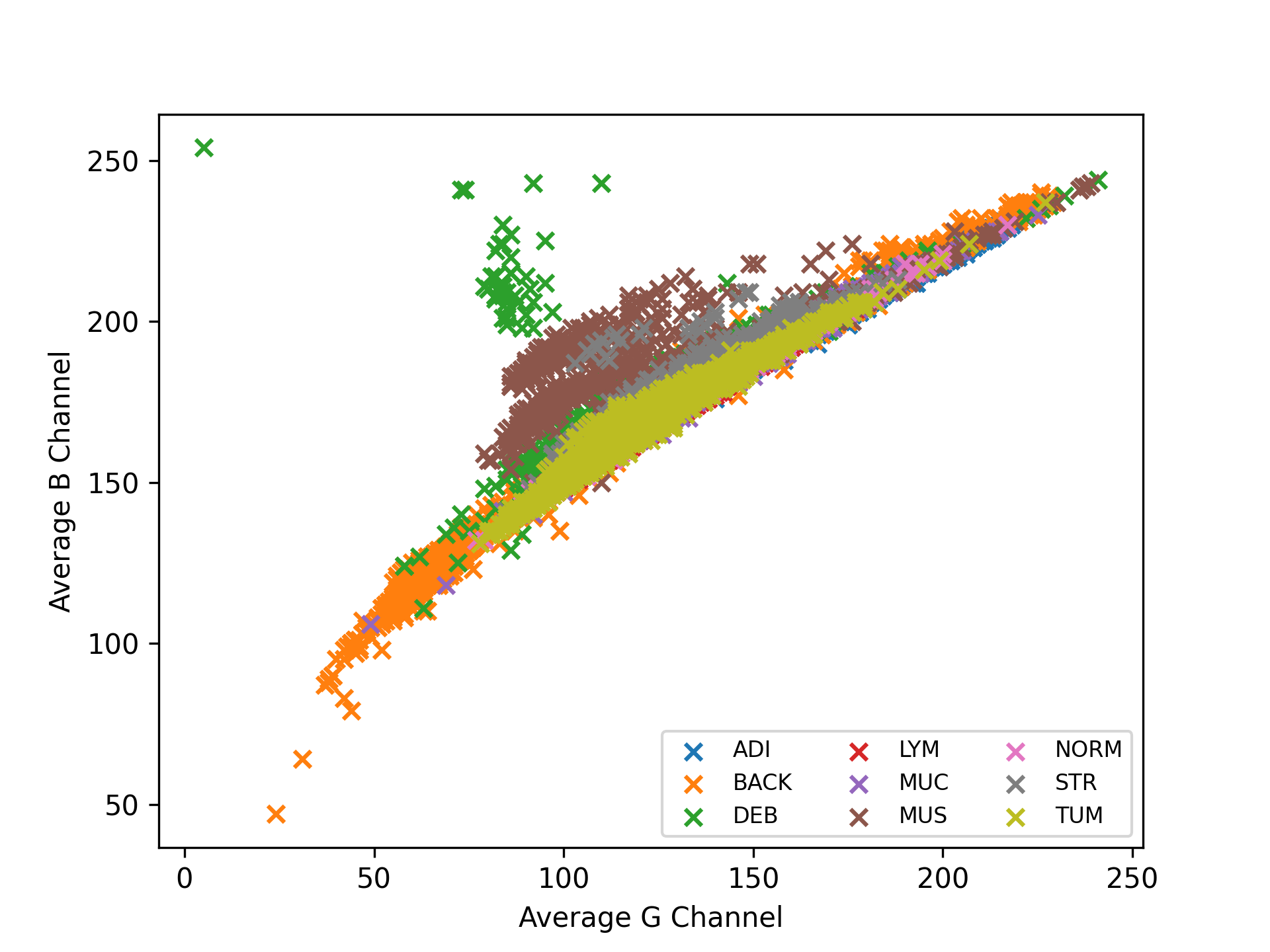}
   \includegraphics[width=0.52\linewidth]{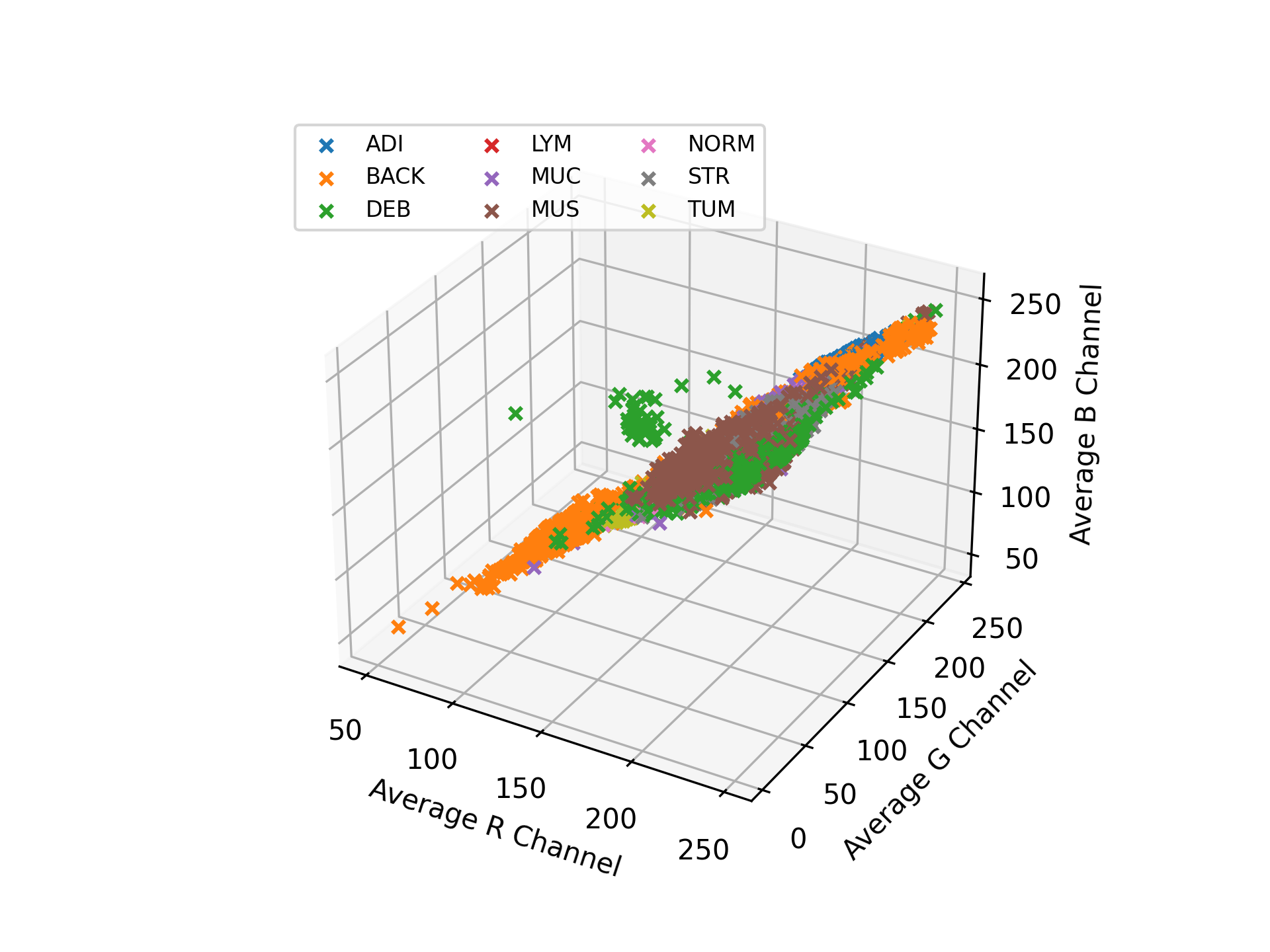}
   \caption{Visualized average red, green and blue color intensities for NCT-CRC-HE training images. Top row shows 2D projections to the corresponding color spaces.}
   \label{fig:dataset_train_rgb}
\end{figure*}

\begin{figure*}[h!]
  \centering
   \includegraphics[width=0.32\linewidth]{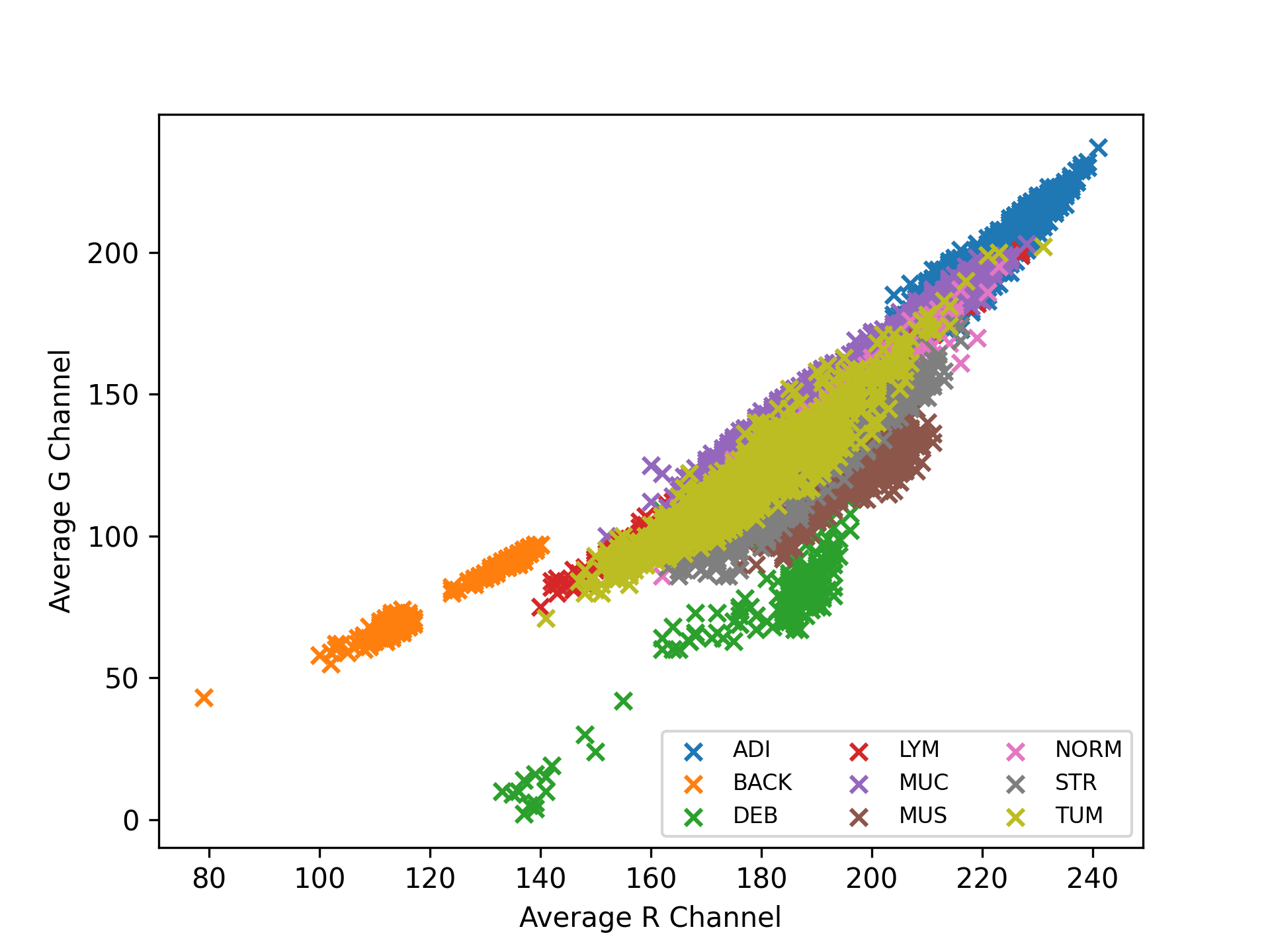}
   \includegraphics[width=0.32\linewidth]{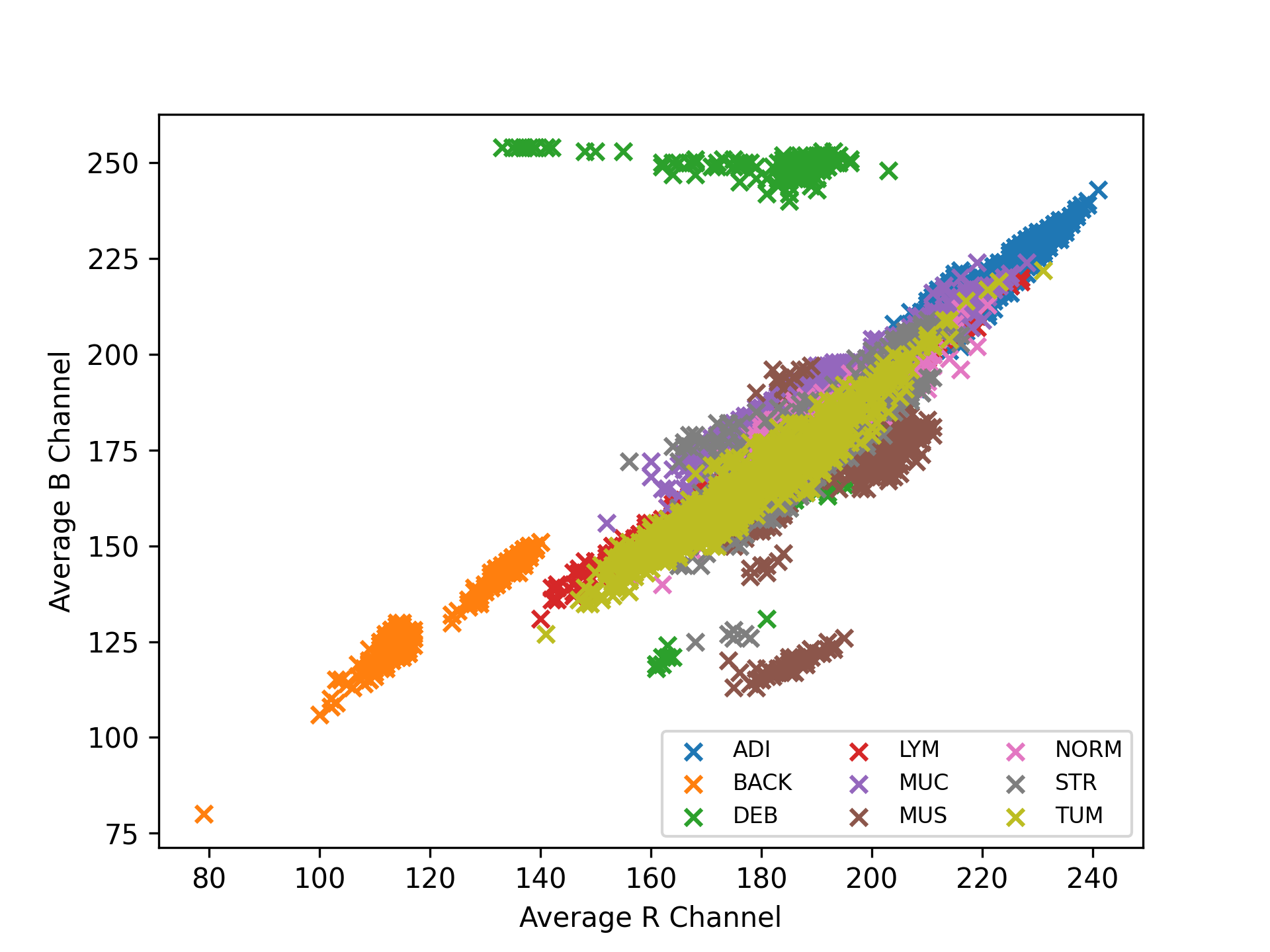}
   \includegraphics[width=0.32\linewidth]{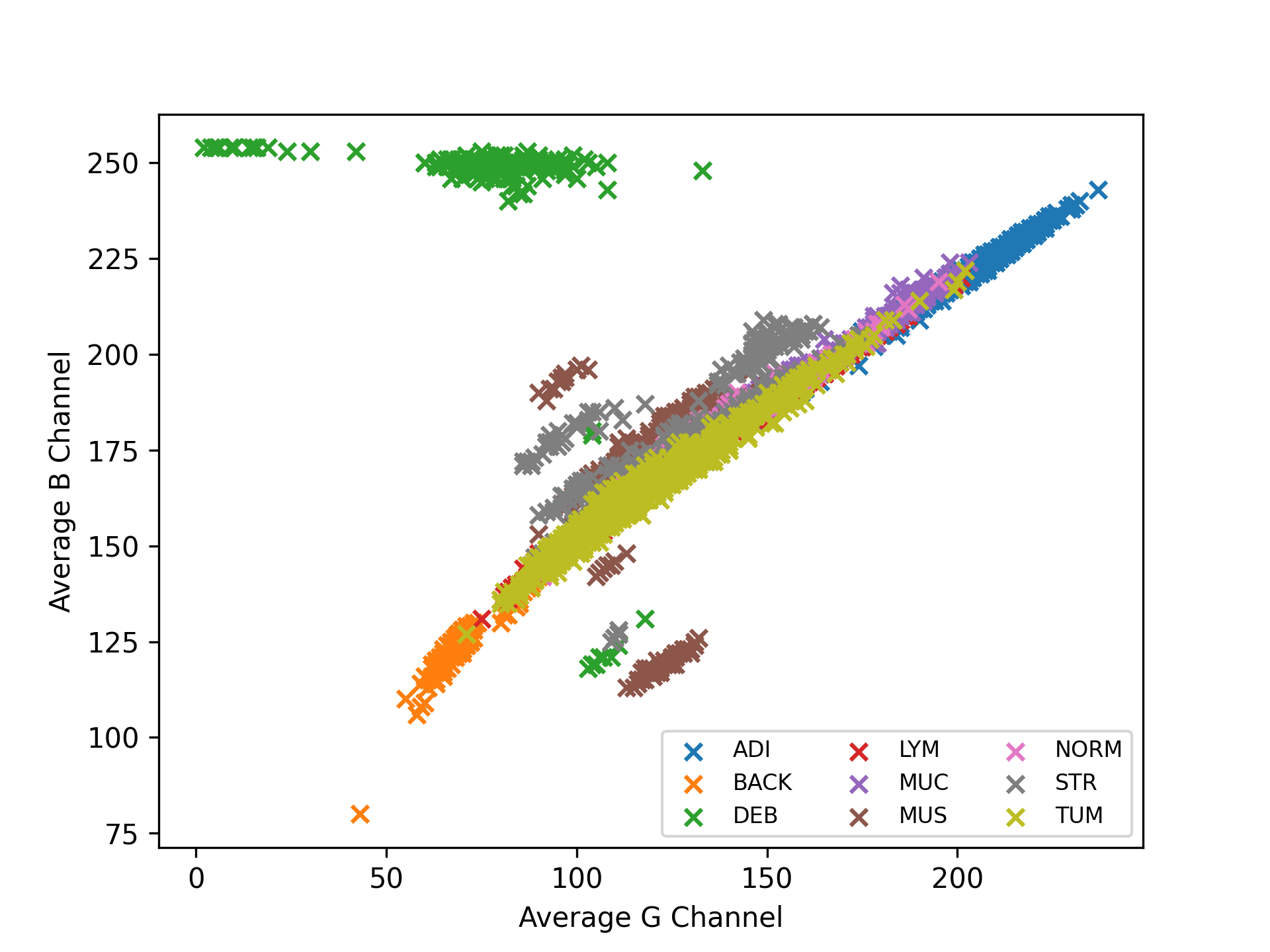}
   \includegraphics[width=0.52\linewidth]{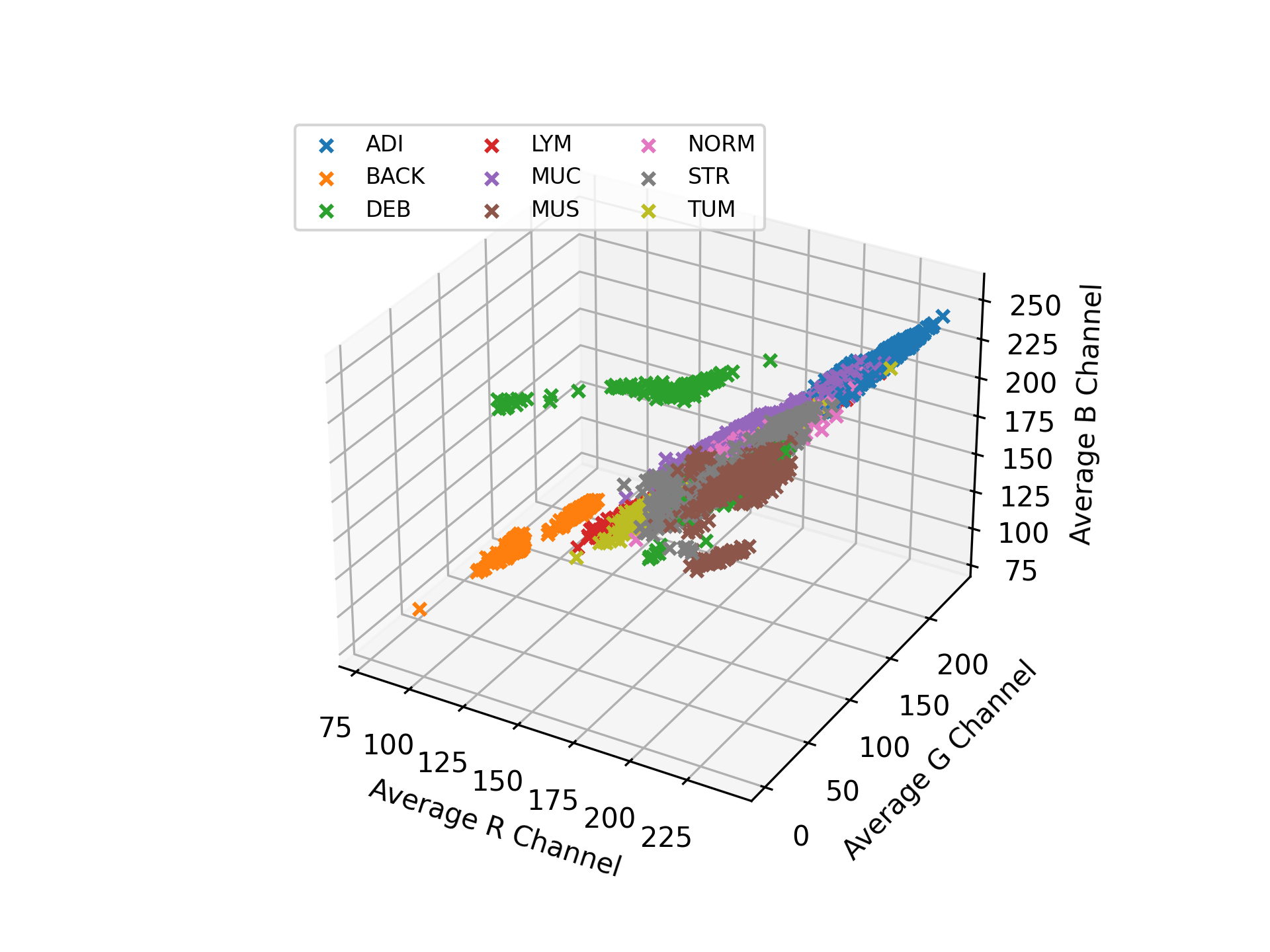}
   \caption{Visualized average red, green and blue color intensities for NCT-CRC-HE test images. Top row shows 2D projections to the corresponding color spaces.}
   \label{fig:dataset_val_rgb}
\end{figure*}

To quantify our observations, we first decided to visualize average red, green and blue color intensities for images from different classes. For this, we averaged the corresponding RGB color channels, thus each image became encoded by three features. The resulting 3D scatter plots as well as 2D projections to the corresponding color spaces are provided in Fig.~\ref{fig:dataset_train_rgb} and Fig.~\ref{fig:dataset_val_rgb} for the training and test sets, respectively. One can observe that samples from different classes are not well mixed, there exists clear overlapping clusters corresponding to different tissue types. Additionally, there is a slight mismatch in RGB intensities distribution between the training and test sets that might potentially contribute to reduced test accuracy for previously proposed transformer and CNN models.

\begin{figure}[t!]
  \centering
   \includegraphics[width=0.6\linewidth]{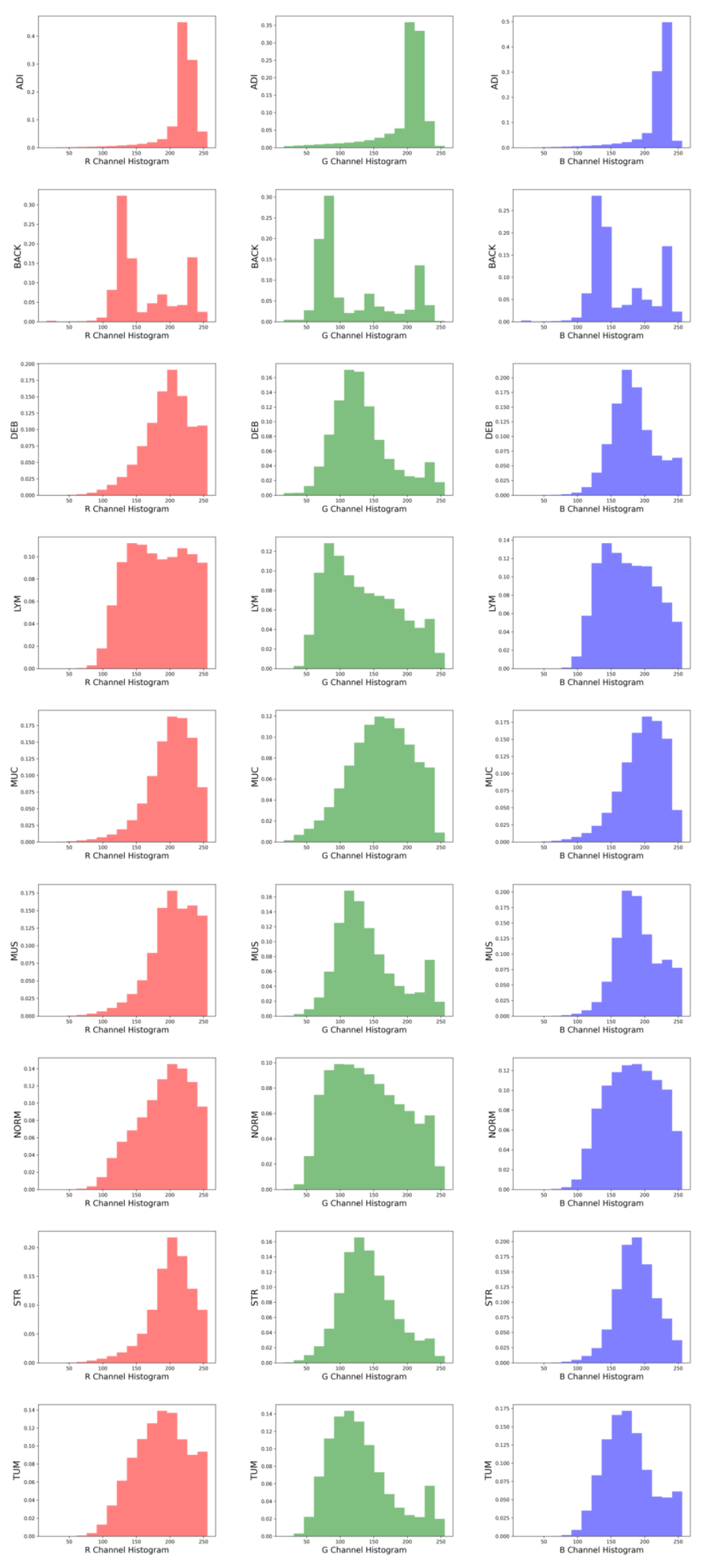}
    %
    \caption{Visualized color histograms for each NCT-CRC-HE tissue class, training set.}
   \vspace{-4mm}
   \label{fig:dataset_train_histograms}
\end{figure}

\begin{figure}[t!]
  \centering
   \includegraphics[width=0.6\linewidth]{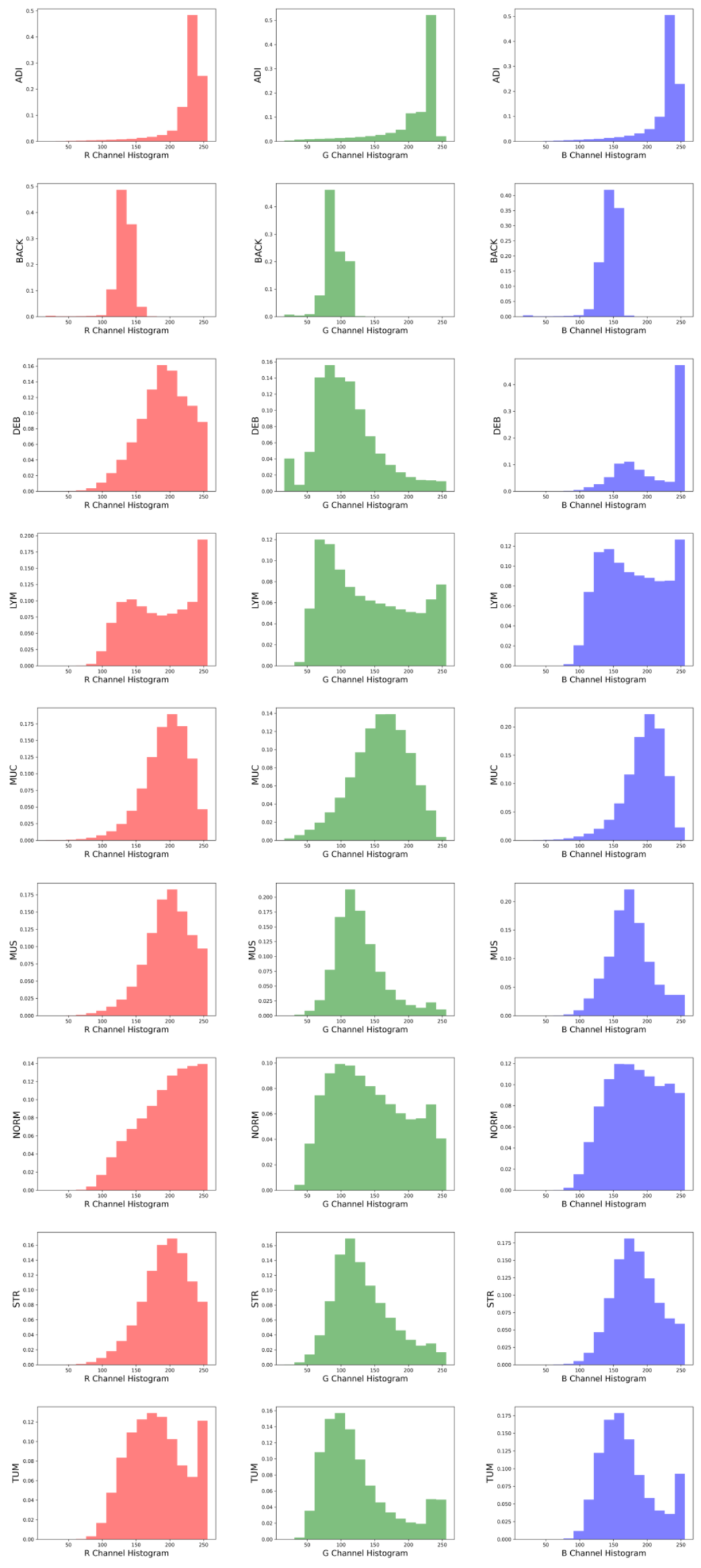}
    %
    \caption{Visualized color histograms for each NCT-CRC-HE tissue class, validation set.}
   \vspace{-4mm}
   \label{fig:dataset_test_histograms}
\end{figure}

Next, we performed a more detailed color distribution analysis by assessing the average color histogram of each class. The results for the training and test NCT-CRC-HE sets are depicted in Fig.~\ref{fig:dataset_train_histograms} and Fig.~\ref{fig:dataset_test_histograms}, respectively. Here, we can see an even better separation of different tissue types: all tissue classes except for \textit{debris (DEB), smooth muscle (MUS)} and \textit{cancer-associated stroma (STR)} have a unique overall histogram profile when combining R, G and B color distributions. This suggests that we can possibly build an accurate classifier for the NCT-CRC-HE dataset by using only color profiles of each image, and not taking into account any complex histopathological features such as cell type composition, vasculature, immune infiltration, \textit{etc}. In the experimental section of this paper, we will validate this assumption by building and evaluating a model which predictions are based only on image histogram data.

We should again highlight a small mismatch in color distributions between the training and test sets. For the latter, there are also noticeable long tails on the right of the histogram for \textit{debris (DEB), lymphocyte (LYM)} and \textit{cancer-associated stroma (TUM)} tissue classes that are caused by ``overexposed'' image regions obtained after color normalization.

\subsection{JPEG Compression Artifacts}

\begin{figure}[t!]
  \centering
  \begin{tabular}{cccccc}
  Adipose & Background & Debris & Normal Mucosa \\
   \includegraphics[width=0.23\linewidth]{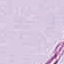} &
   \includegraphics[width=0.23\linewidth]{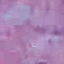} &
   \includegraphics[width=0.23\linewidth]{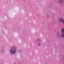} &
   \includegraphics[width=0.23\linewidth]{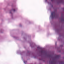} \\
   \includegraphics[width=0.23\linewidth]{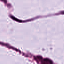} &
   \includegraphics[width=0.23\linewidth]{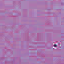} &
   \includegraphics[width=0.23\linewidth]{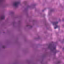} &
   \includegraphics[width=0.23\linewidth]{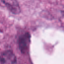} \\
   \includegraphics[width=0.23\linewidth]{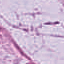} &
   \includegraphics[width=0.23\linewidth]{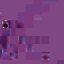} &
   \includegraphics[width=0.23\linewidth]{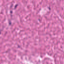} &
   \includegraphics[width=0.23\linewidth]{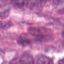} \\
   \includegraphics[width=0.23\linewidth]{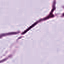} &
   \includegraphics[width=0.23\linewidth]{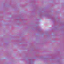} &
   \includegraphics[width=0.23\linewidth]{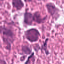} &
   \includegraphics[width=0.23\linewidth]{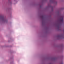} \\
   \end{tabular}
   \caption{Visualization of 64$\times$64 pixel patches extracted from NCT-CRC-HE training images. Severe JPEG compression artifact can be observed on many images of classes \textit{adipose} and \textit{background}, while only minor artifacts are present on images for classes \textit{debris} and \textit{normal mucosa}.}
   \label{fig:dataset_jpeg}
\end{figure}

While all provided images are saved in TIFF format, these are not real raw tissue photos: instead, the compressed JPEG images (obtained presumably after color normalization procedure) were re-saved in this format. The logic behind this action is rather questionable as such procedure only increases the size of the dataset by approximately a factor of 10 without any quality gains. However, a more surprising finding is that the JPEG compression quality level varies across different tissue classes and sometimes even within images of the same class. Figure~\ref{fig:dataset_jpeg} illustrates the observed behavior: \textit{e.g.,} on many images from classes \textit{adipose} and \textit{background} we can see extreme JPEG compression artifacts (checkerboard pattern) corresponding to compression quality level presumably lying between 30-60\%, while for other classes like \textit{debris} and \textit{normal mucosa} this quality level was higher than 70\%. Additionally, for almost all tissues we see intra-class compression quality variation suggesting that different pipelines were used for processing and saving images even of the same class.

This creates a major issue when training deep learning models on such data: as these compression artifacts can be easily detected with just a few convolutional filters, they might become one of the primary features used by the model when learning the decision rule. The contribution of compression artifacts become more significant for larger models that are capable to detect even very small image quality deviations, overfitting to various low-level image properties introduced by WSI pre-processing pipelines.

\subsection{Corrupted Images}

\begin{figure}[t!]
  \centering
  \begin{tabular}{cccccc}
   \includegraphics[width=0.16\linewidth]{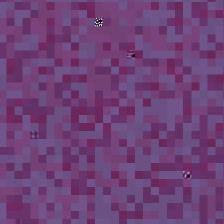} &
   \includegraphics[width=0.16\linewidth]{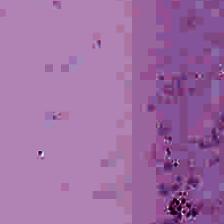} &
   \includegraphics[width=0.16\linewidth]{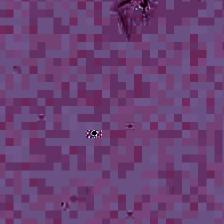} &
   \includegraphics[width=0.16\linewidth]{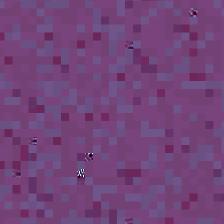} &
   \includegraphics[width=0.16\linewidth]{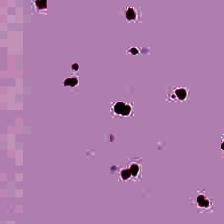} &
   \includegraphics[width=0.16\linewidth]{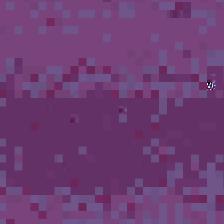} \\
   \includegraphics[width=0.16\linewidth]{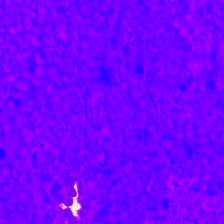} &
   \includegraphics[width=0.16\linewidth]{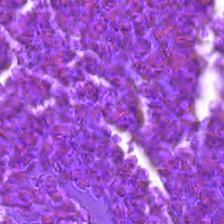} &
   \includegraphics[width=0.16\linewidth]{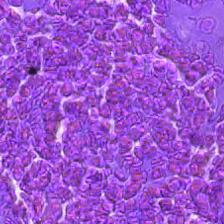} &
   \includegraphics[width=0.16\linewidth]{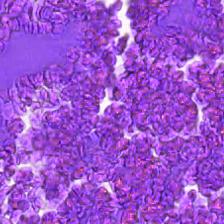} &
   \includegraphics[width=0.16\linewidth]{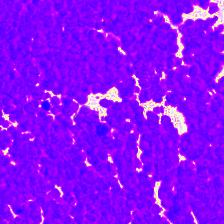} &
   \includegraphics[width=0.16\linewidth]{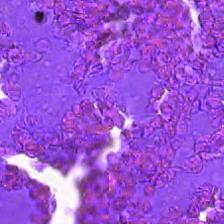} \\
   \end{tabular}
   \caption{Typical corrupted images from class \textit{background} (top row) and \textit{debris} (bottom).}
   \label{fig:dataset_corrupted}
\end{figure}

Visual observation of training and validation patches revealed that the majority of images from class \textit{background} are totally corrupted (Fig.~\ref{fig:dataset_corrupted}, top row): a combination of inappropriately processed image dynamic range obtained after color normalization and extreme JPEG compression rate resulted in pixelated images that no longer represent any biological meaning. While even the simplest machine learning model can correctly classify all images of this class, the resulting accuracy has little relation to the overall task of colorectal cancer tissue analysis.

A similar issue related to incorrect image dynamic range handling can be observed for a fraction of images from class \textit{debris} (Fig.~\ref{fig:dataset_corrupted}, bottom row). Almost half of the test images of this type exhibit over-saturated blue color tint and artificial  looking texture. The origin of this problem can be explained using blue color histogram computed for test images (Fig.~\ref{fig:dataset_test_histograms}, 3rd row): one can see a long tail on the right of the histogram that corresponds to a massive amount of pixels with blue color intensity of 255. We hypothesise that the color normalization procedure for some reason resulted in a shifted dynamic range for the blue color, exceeding the normal maximum pixel intensity value of 255. When the resulting images were saved, all pixels with a higher intensity than 255 were clipped to this value, which resulted in corruptions in image texture and color.

\subsection{Other Potential Issues}

Besides the above mentioned pronounced problems, one can also notice smaller image quality variations related, \textit{e.g.,} to over-sharpening, blur or upsampling that are specific to patches of different tissue types. It was demonstrated in~\cite{fang2023sqad} that deep learning models can uniquely identify the origin of the photo based on such image quality aspects, which potentially allows the network to detect tissue classes without learning tissue morphology. While this should not be generally the case here as there exists more straightforward features allowing to distinguish between different image classes for this dataset, these low-level quality aspects can still introduce some contribution to the final decision rule and accuracy, especially when training large models that tend to learn more complex features.

\section{Proposed Method}
\label{sec:method}

The dataset analysis performed in the previous section led to two important outcomes. First, we identified that the complexity of this specific task itself is relatively low since even basic color information should be sufficient to distinguish between the majority of tissue classes. Secondly, various artifacts and unique low-level image properties specific for different tissue classes might significantly affect model predictions and accuracy, especially since there is a noticeable mismatch in their strength between NCT-CRC-HE training and validation sets.

For the above reasons, we decided to base our solution on a relatively shallow EfficientNet-B0 CNN model \cite{tan2019efficientnet} that has only 4M parameters. Our initial experiments demonstrated that even a slightly larger EfficientNet-B1 network with 6.5M parameters already overfits the data, therefore, unlike all previous solutions that use large network architectures or ensembles of multiple big CNN models, we propose to significantly reduce the model complexity and additionally focus on heavy data augmentation strategy.

The model was initialized with ImageNet weights and trained using the Adam~\cite{kingma2014adam} algorithm with a learning rate of 5e--4 and a weight decay of \mbox{1e--6}. Training data was augmented using random flips, noise, Gaussian blur, color and contrast adjustments. During the inference process, test-time augmentations (averaging the results obtained for the same image flipped vertically and horizontally) were applied to generate the final predictions. The model was trained on one \textit{Nvidia 2070} GPU with 8 GB of vRAM.

\section{Experimental Results}
\label{sec:experiments}

This section provides numerical results obtained with different baseline solutions and the proposed approach based on the EfficientNet-B0 model. We used the conventional NCT-CRC train / validation splits in all experiments, where \textit{NCT-CRC-HE-100K} data is used for training and \textit{CRC-VAL-HE-7K}~--- for validation.

\subsection{Baseline Solution 1: Using R, G and B Color Intensities}

\begin{table}[t!]
\centering
\resizebox{1.0\columnwidth}{!}
{
\begin{tabular}{l|cccccc}
 & Random & \, Avg. R, G, B intensities \, & \, Color Histogram \, & \, ImageNet Features \, & & Ensemble of  \\
Class & \, Classification \, & + Random Forest & + Random Forest & + SVM & \, EfficientNet-B0 \, & \, 2$\times$EfficientNet-B0 \, \\
\hline
\hline
Adipose tissue & 11.0 & 75.2 & 94.2 & 98.3 & 99.3 & 99.6 \\
Background & 12.0 & 99.5 & 100 & 99.5 & 100 & 100 \\
Debris & 10.6 & 68.7 & 57.5 & 94.1 & 98.2 & 99.7 \\
Lymphocytes & 12.5 & 33.6 & 90.2 & 99.2 & 99.7 & 100 \\
Mucus & 10.7 & 44.1 & 92.3 & 96.6 & 99.0 & 99.6 \\
Smooth muscle & 10.6 & 33.8 & 55.2 & 85.3 & 99.2 & 98.3 \\
Normal colon mucosa & 9.6 & 30.5 & 60.5 & 96.0 & 97.6 & 98.1 \\
Cancer-associated stroma & 10.9 & 20.7 & 46.1 & 48.2 & 80.8 & 82.7 \\
Adenocarcinoma epithelium \, & 11.4 & 48.6 & 89.5 & 89.1 & 97.5 & 98.9 \\
\hline
Overall Balanced Accuracy & 11.0 & 50.5 & 76.2 & 89.6 & 96.8  & 97.4 \\
Overall Accuracy & 11.1 & 53.8 & 82.2 & 92.2 & 97.7 & 98.3 \\
\end{tabular}
}
\vspace{2mm}
\caption{Overall and per-class accuracy results for different baseline methods and the proposed EfficientNet-B0 based solution obtained on the CRC-VAL-HE-7K validation set.}
\label{tab:NCT_CRC_per_class}
\vspace{-6mm}
\end{table}

In Section~\ref{sec:nct_exploration} and Fig.~\ref{fig:dataset_train_rgb}, we observed that one might be able to partially separate different tissue classes using only mean red, green and blue color intensities. To validate this assumption, we used these three intensity features generated for all NCT-CRC-HE images and trained a Random Forest classifier model on the obtained data. The results of this experiment are provided in Table~\ref{tab:NCT_CRC_per_class}. While one might expect all tissue classes to be indistinguishable from each other by their mean brightness and intensity values, the considered approach achieved an accuracy of 53.8\%. This means that by using only these three intensity features it is possible to correctly classify more than half of the validation images. This confirms our initial assumption that the majority of the NCT-CRC-HE tissue classes have a unique color signature.


\begin{table}[t!]
\centering
\resizebox{1.0\columnwidth}{!}
{
\begin{tabular}{l|cc}
Method & BA, \% & Accuracy,~\% \\
\hline
\hline
Random Classifier & 11.05 & 11.09 \\
\hline
Average R, G and B color intensities (3 features) + Random Forest & 50.51 & 53.80 \\
Color histogram + Random Forest & 76.17 & 82.20 \\
EfficientNet-B0, ImageNet features + SVM & 89.58 & 92.24 \\
\hline
\hline
DenseNet based solution~\cite{khvostikov2021tissue} & 90.3 & 92.9 \\
VGG19 based solution~\cite{kather2019predicting} & & 94.3 \\
Inception-v3 based solution~\cite{wang2021accurate} & & 94.8 \\
ResNet-50 based solution~\cite{sun2023automatic} & & 94.8 \\
VGG16 based solution~\cite{anju2023finetuned} & & 95.3 \\
CONCH (ViT-Base transformer model)~\cite{lu2023towards}  & 93.0 & -- \\
iBOT (ViT-Large transformer model)~\cite{filiot2023scaling} & 94.4 & 95.8 \\
DINO (ViT transformer model)~\cite{kang2023benchmarking} & 94.5 & 95.9 \\
Ensemble of 4 models \scriptsize{(DenseNet, IncResNetV2, Xception and custom)}\normalsize~\cite{ghosh2021colorectal} \, \, &  & 96.16 \\
Ensemble of 5 models \scriptsize{(Same as~\cite{ghosh2021colorectal} + VGG16)}\normalsize~\cite{kumar2023crccn} &  & 96.26 \\
CTransPath (Swin transformer model)~\cite{wang2022transformer}  &  & 96.52 \\
DeepCMorph (Cell-morphology aware CNN)~\cite{ignatov2024histopathological} & 95.59 & 96.99 \\
\hline
EfficientNet-B0 model & 96.80 & 97.73 \\
Ensemble of 2$\times$EfficientNet-B0 models & \textBF{97.44} & \textBF{98.33} \\
\end{tabular}
}
\vspace{2mm}
\caption{Accuracy results on the CRC-VAL-HE-7K validation set~\cite{kather2019predicting}. BA stands for Balanced Accuracy score.}
\label{tab:NCT_CRC}
\vspace{-6mm}
\end{table}

\subsection{Baseline Solution 2: Using Color Histograms}

Even higher results can be obtained when using more detailed color information extracted from the images. In this experiment, we computed a simple color histogram for each image and for each color channel. The entire 0--255 color intensity range was divided into 16 intervals, which resulted in 48 features generated per image patch. These features were then used by the Random Forest classifier with 200 trees. The results in Table~\ref{tab:NCT_CRC_per_class} demonstrate that this model was able to achieve an overall accuracy of 82.2\% on the entire dataset, and over 89\% of accuracy for five out of nine tissue classes. It should be noted that this model was not using any histopathological features related to cell types and shapes, tissue morphology or immune system activity: only image color distributions largely affected by staining intensities. Despite its high accuracy, this solution has little practical application since its predictions are entirely dependent on the color distribution of the NCT-CRC-HE dataset.

\subsection{Baseline Solution 3: Using ImageNet Features}

One can further improve the results on this dataset without using any specific histopathological information by using ImageNet features. In this experiment, such features were obtained using a pretrained EfficientNet-B0 ImageNet model that generated a feature representation of dimension 1280 for each NCT-CRC-HE image. An SVM classifier was trained on top of these features to learn the decision rule. Table~\ref{tab:NCT_CRC_per_class} presents the results of this solution: the model achieved an accuracy of 92.2\%, for five out of nine tissue classes the accuracy exceeded 96\%. When observing the results of CNN models previously tuned on this dataset (Table~\ref{tab:NCT_CRC}: \textit{DenseNet, VGG19, Inception-V3, ResNet-50}), one can notice that the accuracy improvement does not exceed 3\% compared to this simplistic approach. This suggests that task-specific features that can be learned from this dataset make only a minor contribution to the model's predictive capacity, and the majority of correct decisions can be made based only on simple color and textural information.

\vspace{-1.5mm}
\subsection{EfficientNet-B0 Based Solution}

Next, we performed evaluation of the proposed EfficientNet-B0 based model. We tested two versions of this solution: a single tuned EfficientNet-B0 network and an ensemble of two EfficientNet-B0 models obtained by simple averaging of their predictions. The results of both approaches are shown in Tables~\ref{tab:NCT_CRC_per_class} and~\ref{tab:NCT_CRC}: the proposed solutions achieved an overall accuracy of 97.7\% and 98.3\% for a single model and an ensemble, respectively. With only 4M/8M parameters, they outperformed all previously proposed deep learning models, including foundation transformer-based solutions (CONCH, iBOT, DINO, CTransPath) and a large DeepCMorph model with 87M parameters that was pre-trained to learn cell morphology and tuned on the TCGA dataset with 32 different cancer types. Such results confirm our expectations: due to a low complexity of the task, huge color bias and numerous image artifacts that are not always consistent between the training and validation sets, using large models does not bring any benefits for this dataset. Instead, this might lead to numerous overfitting issues: big models tend to learn complex decision rules, additionally taking into account low-level image quality properties that should not be in general considered in this task. To demonstrate the impact of such low-level image quality aspects on the final model prediction, we performed an extra experiment described below.

\begin{table}[t!]
\centering
\resizebox{1.0\columnwidth}{!}
{
\begin{tabular}{l|c|c|c|c|c|c|c}
Model & \, Base Accuracy \, & \multicolumn{4}{c|}{Texture Deviations: JPEG Artifacts} & \multicolumn{2}{c}{Color Deviations: Hue Alteration} \\
 &  & Quality=80 & Quality=60 & Quality=40 & Quality=20 & -10 / +10 & -20 / +20 \\
\hline
\hline
DeepCMorph~\cite{ignatov2024histopathological} model & 96.99 & \, 96.81 \textBF{(-0.18)} \, & \, 96.23 (-0.76)  \, & \,  95.10 (-1.89)  \, & \,  88.11 (-8.88)  \, & \,  94.96 (-2.03) / 96.46 (-0.51)  \, & \,  91.25 (-5.74) / 92.73 (-4.26) \\
EfficientNet-B0 model & 97.73 & 97.20 (-0.53) & 96.85 (-0.88) & 96.59 (-1.14) & 96.00 (-1.73) & 97.24 (-0.49) / 97.35 \textBF{(-0.38)} & \, 95.67 (-2.06) / 96.36 (-1.37) \\
Ensemble of 2$\times$EfficientNet-B0 \, & 98.33 & \textBF{98.06} (-0.27) & \textBF{97.94 (-0.39)} & \textBF{97.79 (-0.54)} & \textBF{97.59 (-0.74)} & \textBF{98.01 (-0.32)} / \textBF{97.92} (-0.41) & \, \textBF{96.82 (-1.51)} / \textBF{97.30 (-1.03)} \\
\end{tabular}
}
\vspace{2mm}
\caption{The effect of JPEG compression artifacts and color deviations on the DeepCMorph and EfficientNet-B0 classification accuracy estimated on the CRC-VAL-HE-7K validation set.}
\label{tab:NCT_CRC_attack}
\vspace{-8mm}
\end{table}

\vspace{-1.5mm}
\subsection{Estimating the Effect of JPEG Compression Artifacts and Color Bias on Model Predictions}

To analyze how the mentioned compression artifacts and color bias influence the decision rules and model accuracy, we performed an experiment where JPEG artifacts and color alterations were introduced to the images from the validation set and the change in the resulting model classification accuracy was assessed. We used three models: the recently presented DeepCMorph model~\cite{ignatov2024histopathological} as its source codes and pre-trained weights for this dataset are publicly available\,\footnote{\url{https://github.com/aiff22/DeepCMorph}}, the proposed single EfficientNet-B0 model and the ensemble of two EfficientNets. Four different compression quality levels (80\%, 60\%, 40\% and 20\%) and four different color deviation strengths (obtained via image hue alteration by $\pm10$ and $\pm20$) were considered. The results of this experiment are shown in Table~\ref{tab:NCT_CRC_attack}. As hypothesized in the previous section, a significantly bigger DeepCMorph model is considerably more susceptible to both color changes and JPEG artifacts. Severe artifacts (as can be seen on images from classes \textit{adipose} and \textit{background} in Fig.~\ref{fig:dataset_jpeg}) lead to a rapid accuracy drop for this model that reaches 8.8\% for a compression quality level of 20\%. In contrast, both EfficientNet-B0 models show an accuracy decline of only 1.7\% and 0.7\% for a single network and ensemble, respectively, which indicates that JPEG compression artifacts were not used as a main feature when learning the decision rule. A similar situation can be observed in case of color deviations: DeepCMorph model demonstrates a significantly larger accuracy decline even for relatively small color shifts, showing that the color tint of tissue staining should generally have a larger role in its learned decision function.

\section{Conclusion}

In this paper, we deviated from the standard pathway followed by all previous works designing solutions for the NCT-CRC-HE colorectal cancer dataset. As our initial experiments revealed abnormalities in the results and learning curves obtained on this dataset, we started with a detailed exploration of the images it is composed of. The performed dataset analysis revealed a number of critical issues significantly limiting its applicability for designing biomedical tools for histopathological image analysis. The first prominent problem is a strong color signature present for the majority of tissue classes. We demonstrate that by using only three features~-- mean red, green and blue color intensities~-- one can achieve over 50\% of classification accuracy on this dataset. By using a simple color histogram not explicitly capturing histopathological features, it is possible to correctly classify 8 out of 10 test images. In addition to color-related issues, severe JPEG compression artifacts can be found in images belonging to several tissue classes that might contribute to the final decision rules learned by deep learning models. Another problem is related to incorrect dynamic range processing of images obtained after stain normalization, which resulted in a large number of corrupted image patches that, though are easily identifiable even with the simplest machine learning models, no longer have any biological meaning. Taking into account the above issues, we proposed a shallow EfficientNet-B0 based solution that demonstrated an accuracy of over 97.7\% on the CRC-VAL-HE-7K validation set, outperforming all foundation transformer models and cell morphology-aware networks previously proposed for this dataset. Finally, the experiment analyzing the effect of compression artifacts and color bias on deep learning model predictions confirmed that large networks trained on this dataset tend to use low-level image quality aspects for deriving the classification decisions, suggesting that the results obtained on this dataset should be interpreted with caution.

%
%
\bibliographystyle{splncs04}
\bibliography{egbib}
\end{document}